\documentstyle[prl,twocolumn,aps]{revtex}
\begin{document}
\draft

\title{
Zero-bias anomalies and boson-assisted tunneling through quantum dots}

\author{J\"urgen K\"onig$^1$, Herbert Schoeller$^{1,2}$
and Gerd Sch\"on$^1$}

\address{
$^1$ Institut f\"ur Theoretische Festk\"orperphysik, Universit\"at
Karlsruhe, 76128 Karlsruhe, Germany\\
$^2$ Department of Physics, Simon Fraser University, Burnaby, B.C.,
V5A 1S6, Canada}

\date{\today}

\maketitle

\begin{abstract}
We study resonant tunneling through a
quantum dot with one degenerate level in the presence of a
strong Coulomb repulsion and a bosonic environment.
Using a real-time approach we calculate the spectral density and the
nonlinear current within a conserving approximation.
The spectral density shows a multiplet of Kondo peaks
split by the transport voltage and boson frequencies.
As a consequence we find a zero-bias
anomaly in the differential conductance
which can show a local maximum or minimum depending on the
level position.
The results are compared with recent experiments.

\end{abstract}

Transport phenomena through discrete energy levels
in quantum dots have been studied
by perturbation theory \cite{Ave-Kor-Lik,Bru-HS}
and beyond \cite{Gro1,Mei-Win-Lee,Het-HS}.
In general, resonant tunneling phenomena
and Kondo effects in nonequilibrium become important,
which have been measured recently by Ralph \& Buhrman
\cite{Ral-Buh1}.
In metallic islands, the Coulomb blockade is strongly influenced by
inelastic  interactions with bosonic degrees of freedom, such as
fluctuations of the electrodynamic environment \cite{Ing-Naz}
or applied time-dependent fields \cite{Kou-etal1}.
The study of inelastic interactions in quantum dots with few levels
has started only recently, either for the nondegenerate case
\cite{Win-Jac-Wil,Ima-Pon-Ave} or more general,
in the presence of time-dependent fields and Coulomb blockade
\cite{Bru-HS,Het-HS}. In earlier work we have studied
the influence of bosonic
fields in the nonequilibrium Anderson model
in the perturbative regime  \cite{Sta} and found resonant side
peaks in the Coulomb oscillations.

The purpose of the present letter is to investigate the influence of
external quantum-mechanical
fields on transport phenomena through ultrasmall quantum dots at
low temperatures and frequencies (compared to the intrinsic broadening
of the resonant state in the dot).
This requires a description of the Kondo effect,
generalized to nonequilibrium situations and including coupling
to bosonic fields.
For the nonperturbative treatment of the tunneling we
apply a  real-time, nonequilibrium
many-body approach developed recently
\cite{Schoell-Schoen,Koe-Schoell-Schoen}
to a quantum dot with one level and spin degeneracy $M$.
For $M \ge 2$ and low lying dot level $\epsilon$
we obtain the usual Kondo peaks
at the Fermi levels $\mu_\alpha$ of the reservoirs \cite{Mei-Win-Lee}.
However, the emission of
bosons causes additional Kondo singularities,
for a one mode field
at $\mu_\alpha+n\omega_B$ ($n=\pm 1,\pm 2,\ldots$).

Furthermore, we will analyze the effect of the
singularities in the spectral density on the differential
conductance as function of the bias voltage.
For a low lying level we obtain the
well-known zero bias {\it maximum} \cite{Mei-Win-Lee,Het-HS,Ral-Buh1},
whereas for a level close to the chemical potentials of the
reservoirs we find a zero bias {\it minimum}.
The coupling to bosons gives rise to
satellite anomalies, which can be traced back to
the corresponding satellite peaks in the spectral density.
In a certain range of gate voltages, for $M=2$ and
in the absence of bosons, we
find that the temperature and bias voltage dependence of the conductance
coincides with recent measurements of zero-bias minima in
point-contacts \cite{Ral-Buh2}. Therefore, in addition to
Refs.~\cite{Ral-Lud-Del-Buh,Het-Kro-Her,Win-Alt-Mei}, we
propose here another possible interpretation of this
experiment.

We consider a dot containing only one energy level with
degeneracy $M$ connected via high tunnel barriers to
reservoirs of noninteracting electrons.
We, furthermore, include a coupling to bosonic modes
representing phonons, photons or
fluctuations of the electrodynamic environment.
Our model Hamiltonian reads
$H=H_0+H_T$, where $H_0$ describes the decoupled system and
$H_T$ the tunneling between leads and dot.
We write $H_0=H_R+H_D$ where
$H_R=\sum_{k\sigma\alpha}\epsilon_{k\alpha}
a^\dagger_{k\sigma\alpha}a_{k\sigma\alpha}$
refers to the reservoirs ($\sigma$ and
$\alpha$ are spin and reservoir indices). Furthermore,
($\hbar=k_B=1$)
\begin{eqnarray}\nonumber
	H_D=&\epsilon_0\hat{N}+U_0\sum_{\sigma < \sigma^\prime}
	n_\sigma n_{\sigma^\prime}\\
&+\sum_q \omega_q d_q^\dagger d_q
	+\hat{N}\sum_q g_q (d_q + d_q^\dagger)
\label{1}
\end{eqnarray}
describes the isolated dot with $M$ spin degenerate levels
at position $\epsilon_0$,
Coulomb repulsion $U_0$, bosonic modes $\omega_q$ and
electron-boson coupling $g_q$.
The number of particles on the dot with spin $\sigma$ is denoted by
$n_\sigma=c^\dagger_\sigma c_\sigma$, and
$\hat{N} = \sum_{\sigma} n_{\sigma}$. Finally, the
tunneling term is given by $H_T=\sum_{k\sigma\alpha}
(T^\alpha_{k} a^\dagger_{k\sigma\alpha}c_\sigma + h.c.)$.

This Hamiltonian can be rewritten
after a unitary transformation \cite{Mah} defined by
$V=\exp(-i\hat{N}\varphi)$ and $\varphi=i\sum_q
(g_q/\omega_q)(d_q^\dagger-d_q)$.
We get $\bar{H}=VHV^{-1}=\bar{H}_0+\bar{H}_T$,
where $\bar{H}_0=H_R+\bar{H}_D$,
$\bar{H}_D=\epsilon\hat{N}
+U\sum_{\sigma<\sigma^\prime}n_\sigma n_{\sigma^\prime}
+\sum_q \omega_q d^\dagger_q d_q$ and
$\bar{H}_T=\sum_{k\sigma\alpha}(T^\alpha_{k} a^\dagger_{k\sigma
\alpha}c_\sigma e^{i\varphi}+h.c.)$.
Due to the electron-boson interaction the level position and the
Coulomb repulsion are renormalized, $\epsilon=\epsilon_0-\sum_q
{g_q^2/\omega_q}$ and $U=U_0-2\sum_q{g_q^2/\omega_q}$, and
the tunneling part contains now phase factors $e^{\pm i\varphi}$.

In lowest order perturbation theory the rates for tunneling
in and out of the dot to reservoir $\alpha$ are
\begin{equation}\label{3}
	\gamma_\alpha^\pm(E)=\int dE^\prime \bar{\gamma}^\pm_\alpha (E^\prime)
	P^\pm (E-E^\prime) \, ,
\end{equation}
where $\bar{\gamma}_\alpha^\pm(E)={1/(2\pi)} \Gamma_\alpha(E)f^\pm_
\alpha(E)$ is the classical rate without bosons,
$\Gamma_\alpha (E)=2\pi\sum_k |T^\alpha_k|^2 \delta (E-
\epsilon_{k\alpha})$, and $f_\alpha^+(E)$ is
the Fermi distribution of reservoir $\alpha$ with
chemical potential $\mu_\alpha$ while $f_\alpha^-(E)=1-f^+_\alpha (E)$.
Furthermore,
$P^\pm (E)$ describes the probability for an electron to absorb (for $P^+$)
or emit (for $P^-$) the boson energy E. It is \cite{Ing-Naz}
\begin{equation}\label{4}
P^\pm (E)={1\over 2\pi}\int dt e^{iEt} <e^{i\varphi (0)}
e^{-i\varphi (\pm t)}>_0
\end{equation}
where $<...>_0$ denotes the
expectation value with respect to the free boson Hamiltonian.
The classical rates together with a master equation
are sufficient in the perturbative regime $\Gamma
= \sum_{\alpha} \Gamma_{\alpha} \ll T$ \cite{Sta}.
In this letter we are interested in
temperatures and frequencies which are of the order or smaller
than the intrinsic level broadening $\Gamma$, which
requires a nonperturbative treatment in $\Gamma$.
To achieve this we  use
a real-time technique developed in
\cite{Schoell-Schoen,Koe-Schoell-Schoen} which provides a
natural generalization of the classical and cotunneling
theory to the physics of resonant tunneling. For details
we refer to these papers. Here we only
sketch the derivation and quote the results.

We develop a diagrammatic approach by
expanding in the tunneling Hamiltonian $\bar{H}_T$.
Since $\bar{H}_0$ contains interaction
terms, this can not be done by usual Green's function techniques
since Wick's theorem does not apply. However,
we can use it with respect to the field operators of the
reservoirs, since $\bar{H}_0$ is bilinear in these operators.
As an example let us consider the reduced density operator of the
dot. We assume that the reservoirs and the boson bath
remain in thermal equilibrium. On the other hand, we
want to study the nonequilibrium time evolution of the dot.
An effective description in terms of the dot degrees of freedom
can be derived by expanding all propagators in $\bar{H}_T$
and tracing out the reservoirs by applying Wick's theorem
for them. A matrix element of the reduced density
operator can then be visualized as shown in Fig.(\ref{fig1}).
The forward and the backward propagator (Keldysh contour) are coupled
by ``tunneling lines'' associated with the junctions to each
reservoir $\alpha$.
Each tunneling line with energy $E$ represents
the rate $\bar{\gamma}_\alpha^+(E)$ if the
line is directed backward with respect to the closed time path and
$\bar{\gamma}_\alpha^-(E)$ if it is directed forward.
Because of Fermi-Dirac statistics, we get a factor $-1$ if two tunneling
lines cross each other.
The tunneling lines are associated with changes of the
state of the dot, as indicated on the closed time path.
For large Coulomb repulsion $U$ we can restrict ourselves
 to states with $N=0,1$.
The coupling to the bosons is introduced by connecting all
vertices in all possible ways by boson lines with a certain
energy $E$ (the direction can be chosen in an arbitrary way).
The rule for the contribution of the boson lines is exactly
the same as for the reservoir lines except that we have to
replace $\bar{\gamma}^\pm_\alpha$ by $P^\pm$.
Finally, as in the case for metallic islands \cite{Schoell-Schoen},
we have to associate to each tunneling vertex at time $t$
on the contour a factor $\exp{i\Delta E\,t}$
where $\Delta E$ is the difference of the out- and
incoming energies.
If the vertex lies on the backward propagator it
acquires a factor $-1$. Analogous graphical rules hold for
the Green's functions of the dot as well, the only difference is the
occurrence of external vertices.

In leading order, we include only boson lines between vertices
which are already connected by tunneling lines.
This simply amounts to a dressing of the tunneling
lines $\bar{\gamma} \rightarrow \gamma$, and
the diagrams look identical to those without bosons.
The approximation, while neglecting many diagrams,
describes well the
spectral density of the dot at resonance points.
The reason is that position and value of the peaks of the
spectral density are determined by a self-energy
$\sigma$ (see Eq.~(\ref{5})) which is calculated here in lowest order
perturbation theory in $\Gamma$ including the bosons. Higher orders
are small for high tunnel barriers.

Similar to the case of metallic islands
\cite{Schoell-Schoen,Koe-Schoell-Schoen} we proceed in
a conserving approximation, which takes into account
non-diagonal matrix elements of the total density matrix
up to the difference of one electron-hole pair excitation in
the reservoirs.  The difference in the case of a single
level quantum dot is that we have
now $M$ possibilities for the occupied state.
The analytic resummation of the corresponding diagrams yields
for the transitions between $N=0(1)$ and $1(0)$ the rates
$\Sigma^\pm =\lambda \int dE\,\gamma^\pm (E) |R(E)|^2$.
Here $R(E)=[E-\epsilon-\sigma(E)]^{-1}$ defines a resolvent with broadening
and energy renormalization given by the self-energy
\begin{equation}\label{5}
	\sigma(E)=\int dE^\prime {M \gamma^+ (E^\prime)
	+\gamma^-(E^\prime)\over E-E^\prime+i0^+}
\end{equation}
where $\gamma^\pm=\sum_\alpha\gamma_\alpha^\pm$
and $\lambda^{-1}=\int dE |R(E)|^2$.

In the classical limit $\Gamma << T$ we recover for $\Sigma^\pm$
the classical rates $\gamma^\pm$.
The stationary probabilities $P_0$ and $P_1$ for an unoccupied or
occupied dot state follow from the kinetic equation which uses
the rates as input $P_1\Sigma^- -P_0 M\Sigma^+=0$. Together
with $P_0+P_1=1$ we obtain
$P_0=\int dE \gamma^-(E)|R(E)|^2$ and
$P_1=M\int dE \gamma^+(E)|R(E)|^2$.

Summing equivalent diagrams for the real-time Green's functions
of the dot we obtain the spectral density $\rho \equiv (G^< - G^>)/(2\pi i)$
\begin{equation}\label{7}
\rho (E) =
\int dE^\prime\sum_{r=\pm}\gamma^r(E^\prime)
P^{-r} (E^\prime-E)|R(E^\prime)|^2
\end{equation}
and for the current $I_\alpha$ flowing into reservoir $\alpha$
\begin{equation}\label{8}
I_\alpha = e2\pi M\sum_{\alpha^\prime}\int dE^\prime
\sum_{r=\pm} r\gamma^{-r}_\alpha (E^\prime)
\gamma^r_{\alpha^\prime}(E^\prime)|R(E^\prime)|^2.
\end{equation}
For the special case of two reservoirs $\alpha=L/R$
and constant level broadening $\Gamma=\Gamma_L=\Gamma_R$
the current $I=I_L=-I_R$ can be written as
\begin{equation}\label{9}
I=e{M\over 2}\Gamma\int dE\rho (E) [f^+_R (E)-f^+_L (E)] \;.
\end{equation}
Our results satisfy all sum rules
together with current conservation, and one can
prove particle-hole symmetry in the case $M=1$.

The difference to other approaches in the $M=1$ case
\cite{Win-Jac-Wil,Ima-Pon-Ave} is clearly
displayed by the effect of the self-energy $\sigma(E)$ which
determines via the resolvent $R(E)$ the position of the maxima of
the spectral density (\ref{7}). In all previous works,
$\sigma(E)$ has been approximated by a
constant. We find that the energy dependence
of $\sigma(E)$ cannot been neglected if the
temperature $T$ and the typical frequency $\omega_B$ of the
bosons are smaller than $\Gamma$. To derive this analytically
we consider from now on a one-mode environment (Einstein
model) with boson frequency $\omega_q=\omega_B$.
Defining $g=\sum_q {g_q^2 / \omega_B^2}$ we obtain
$P^\pm(E)=\sum_n p_n \delta(E\pm n\omega_B)$,
where $p_n=e^{-g(1+2N_0)}e^{n\omega_B/2T_B}
I_n(2gN_0 e^{\omega_B/2T_B})$ is the probability
for the emission of n bosons with frequency $\omega_B$.
Here, $N_0$ is the Bose function, $I_n$ denotes
the modified Bessel function. The temperature
of the boson bath is $T_B$. In real experiments
it can be different from the electron temperature $T$.
Using (\ref{3}) and (\ref{5}) we obtain
\begin{eqnarray}\nonumber
Re\,\sigma(E)=\sum_{n,\alpha} (M p_n - p_{-n}){\Gamma\over 2\pi}
            \Big{[}  ln \left( {E_C\over 2\pi T} \right)   \\
	- Re\,\Psi \left({1\over 2}
    -i{E+n\omega_B-\mu_\alpha \over 2\pi T} \right) \Big{]}
\label{12}
\end{eqnarray}
and $Im\, \sigma(E)=-\pi\sum_n p_n[M \bar{\gamma}^+(E+n\omega_B)+
\bar{\gamma}^-(E-n\omega_B)]$.
Here $\Psi$ denotes the digamma function, and we have chosen
in the energy integrals a Lorentzian
cut-off at $E_C$.
The real part of $\sigma$ depends logarithmically on energy,
temperature, voltage and frequency. These logarithmic terms
are typical for the occurrence of Kondo peaks and do
not cancel for $M\ge 2$ or $p_n\neq p_{-n}$. Hence
we anticipate logarithmic singularities
not only for the degenerate case but also for a single
dot level without spin since the probabilities for absorption and
emission of bosons are different.
This is an important difference
to the case of classical time-dependent fields
\cite{Het-HS} where both probabilities are equal.

Fig.~(\ref{fig2}) shows a typical series of pictures for the
spectral density at different voltages for a low lying level
$\epsilon$. Without bias and
$M=2$, we obtain the usual Kondo peak near the Fermi level
(which we choose as zero energy).
Due to emission of bosons, there are now additional
resonances at multiples of $\omega_B$.
For finite bias voltage, all peaks split and decrease in
magnitude.

The resonances in the spectral density can be seen most
pronounced in the nonlinear differential conductance
as function of the bias voltage $V$.
Fig.~(\ref{fig3}) shows the differential
conductance for a low lying level $\epsilon$.
As usual we find a zero-bias maximum \cite{Mei-Win-Lee,Het-HS,Ral-Buh1}
since the splitting of the Kondo peak leads to an
overall decrease of the spectral density in the
energy range $|E|<eV$ (see inset of Fig.~(\ref{fig3})).
Due to emission of bosons we observe also
a set of symmetric satellite maxima. They can be
traced back to the fact that pairs of Kondo peaks
can merge if the bias voltage is given by
multiples of the boson frequency (see Fig.~(\ref{fig2})). This
gives rise to pronounced Kondo peaks at $E=\pm eV/2$ and thus
to an increase of the spectral density with bias voltage near
these points.

The differential conductance for $\epsilon$ near zero energy
is shown in
Fig.~(\ref{fig4}) with and without bosons. Surprisingly we
find that the whole structure is inverted compared to the
$\epsilon<0$ case and we find a zero-bias anomaly
although the Kondo peak at zero energy is absent.
The contributions of sequential and
cotunneling lead only to an overall shift of the
differential conductance without any interesting structure.
This shows clearly
that the influence of the logarithmic terms in
$\sigma(E)$ is still important. They lead to an
overall increase of the spectral density near zero
energy with bias voltage. In the presence
of bosons we obtain satellite steps at $|eV|=m\omega_B$.

The occurrence of zero-bias minima is
well known for Kondo scattering from magnetic impurities
\cite{Jan-etal}. Here we have shown that zero-bias
minima can also occur by resonant tunneling via local
impurities if the level position is
high enough to enter the mixed valence regime. We have
also compared the scaling behavior of the conductance
as function of temperature and bias voltage
with recent experiments of Ralph \& Buhrman
\cite{Ral-Buh2} (see insets of Fig.~(\ref{fig4})).
The coincidence is quite remarkable.
The explanation of this experiment,
 either interpreting it as 2-channel
Kondo scattering from atomic tunneling systems
\cite{Ral-Lud-Del-Buh,Het-Kro-Her} or by tunneling
into a disordered metal \cite{Win-Alt-Mei},
is still controversial. The mechanism
described in this work offers another possibility although
the magnetic field dependence of the experiments remains unexplained.

Finally, we also investigate the differential conductance
at fixed bias voltage as function of the position of the
dot level, which experimentally can be varied by changing
the gate voltage coupled capacitively to the dot.
Fig.~(\ref{fig5}) shows the classically expected pair of peaks at
$|\epsilon|=eV/2$ together with satellites between
the main peaks (due to emission and absorption)
and peaks for $|\epsilon|>eV/2$ (only due to absorption
of bosons).
The imaginary part of $\sigma(E)$ gives rise to a classically
unexpected asymmetry of the peak heights. The peak at $\epsilon
=eV/2$ is larger than the one at $\epsilon=-eV/2$ since
$|Im\, \sigma(E)|=\pi |M\gamma^+(E)+\gamma^-(E)|$ is always smaller
for higher energies (except for the $M=1$ case
where particle-hole symmetry holds). This demonstrates a
significant effect due to the broadening of the spectral
density by quantum fluctuations.

In conclusion, we have studied for the first time
low-temperature transport in the nonequilibrium Anderson
model with bosonic interactions. A one-mode
environment yields new Kondo resonances in the
spectral density which can be probed by the measurement
of the nonlinear differential conductance. We have shown
that both the gate and bias voltage dependence is
important. Quantum fluctuations due to resonant tunneling
yield zero-bias anomalies as function of the bias voltage,
which can be changed from maxima to minima by varying the
gate voltage. We found similarities to recent experiments.

We like to thank D. Averin, J. von Delft and M. Hettler
for useful discussions. Our work was
supported by the ``Deutsche Forschungsgemeinschaft''
as part of ``SFB 195'' and
the Swiss National Science Foundation
(H.S.).

\begin{figure}
\vspace{0.3cm}
\caption{A diagram showing various tunneling processes:
	sequential tunneling in the left and right junctions,
	a term preserving the norm, a cotunneling process,
	and resonant tunneling.}
\label{fig1}
\end{figure}
\begin{figure}
\caption{The spectral density for $T=T_B=0.01\Gamma$, $\epsilon=-4\Gamma$,
         $g=0.2$, $\omega_B=0.5\Gamma$ and $E_C=100\Gamma$ at different
         voltages.
         For $V=0$ there are resonances at multiples of $\omega_B$, which split
         for finite bias voltage.}
\label{fig2}
\end{figure}
\begin{figure}
\caption{The differential conductance vs. bias voltage for
         $T=T_B=0.01\Gamma$, $\epsilon=-4\Gamma$, $\omega_B=0.5\Gamma$ and
         $E_C=100\Gamma$. The curves show a maximum at zero bias and
         satellite maxima at multiples of $\omega_B$ for a finite
         electron-boson coupling. Inset ($g=0$): increasing voltage leads to an
         overall decrease of the spectral density in the range $|E|<eV$,
         which explains the zero-bias maximum.}
\label{fig3}
\end{figure}
\begin{figure}
\caption[A]
         {The differential conductance vs. bias voltage for $T=T_B=0.01\Gamma$,
         $\epsilon=0$,
         $\omega_B=0.5\Gamma$ and $E_C=100\Gamma$. The curves show a minimum
         at zero bias and steps at multiples of $\omega_B$ for a finite
         electron-boson coupling. Left inset: the rescaled curves for $g=0$ at
         different temperatures collapse onto one curve. Right inset: The
         temperature dependence of the linear conductance (solid line)
         coincides with experimental data from \cite{Ral-Buh2} (triangles).}
\label{fig4}
\end{figure}
\begin{figure}
\caption{The differential conductance as a function of $\epsilon$ for
         $T=0.25\Gamma$, $eV=30\Gamma$, $g=0.3$, $\omega_B=5\Gamma$ and
         $E_C=500\Gamma$.}
\label{fig5}
\end{figure}


\begin{references}

\bibitem{Ave-Kor-Lik}
D.V. Averin, A.N. Korotkov, and K.K. Likharev, Phys. Rev. {\bf B44},
6199 (1991);
C.W.J. Beenakker, Phys. Rev. {\bf B44}, 1646 (1991);
L.I. Glazman and K.A. Matveev, JETP Lett {\bf 48}, 445 (1988).

\bibitem{Bru-HS}
C. Bruder and H. Schoeller, Phys. Rev. Lett. {\bf 72}, 1076 (1994).

\bibitem{Gro1}
T.K. Ng and P.A. Lee, Phys. Rev. Lett. {\bf 61}, 1768 (1988);
L.I. Glazman and M.E. Raikh, JETP Lett. {\bf 47}, 452 (1988);
S. Hershfield, J.H. Davies, and J.W. Wilkins,
Phys. Rev. Lett. {\bf 67}, 3720 (1991)

\bibitem{Mei-Win-Lee}
Y. Meir, N.S. Wingreen, and P.A. Lee, Phys. Rev. Lett.
{\bf 70}, 2601 (1993); N.S. Wingreen and Y. Meir, Phys.
Rev. {\bf B49}, 11040 (1994).

\bibitem{Het-HS}
M.H. Hettler and H. Schoeller, to appear in Phys. Rev. Lett.

\bibitem{Ral-Buh1}
D.C. Ralph and R.A. Buhrman, Phys. Rev. Lett. {\bf 72}, 3401 (1994).

\bibitem{Ing-Naz}
M.H. Devoret et al., Phys. Rev. Lett. {\bf 64}, 1824 (1990);
S.M. Girvin et al. Phys. Rev. Lett. {\bf 64}, 3183 (1990);
A.A. Odintsov, Sov. Phys. JETP {\bf 67}, 1265 (1988);
A.A. Odintsov, V. Bubanja, and G. Sch\"on, Phys. Rev.
{\bf B46}, 6875 (1992);
K. Flensberg {\it et al.}, Phys. Scripta {\bf T42}, 189 (1992).

\bibitem{Kou-etal1}
L.P. Kouwenhoven {\it et al.}, Phys. Rev. {\bf B50}, 2019 (1994);
I.A. Devyatov and K.K. Likharev, Physica {\bf B194-196},
1341 (1994);

\bibitem{Win-Jac-Wil}
N.S. Wingreen, K.W. Jacobsen, and J.W. Wilkins,
Phys. Rev. Lett. {\bf 61}, 1396 (1988);
L.I. Glazman and R.I. Shekhter, Sov. Phys. JETP {\bf 67}, 163 (1988);
M. Jonson, Phys. Rev. {\bf B39}, 5924 (1989).

\bibitem{Ima-Pon-Ave}
H.T. Imam, V.V. Ponomarenko, and D.V. Averin,
Phys. Rev {\bf B50}, 18288 (1994).

\bibitem{Sta}
J. Stampe, H. Schoeller, unpublished.

\bibitem{Schoell-Schoen}
H. Schoeller and G. Sch\"on, Phys. Rev. {\bf B50},
18436 (1994); Physica {\bf B203}, 423 (1994).

\bibitem{Koe-Schoell-Schoen}
J. K\"onig, H. Schoeller and G. Sch\"on, to be published in
Europhys. Lett.; J. K\"onig, H. Schoeller, G. Sch\"on
and R. Fazio, in {\it Quantum Dynamics of Submicron Structures}, eds.
H. A. Cerdeira et al., NATO ASI Series E, Vol. 291
(Kluwer, 1995), p.221

\bibitem{Ral-Buh2}
D.C. Ralph and R.A. Buhrman, Phys. Rev. Lett. {\bf 69},
2118 (1992); Phys. Rev. {\bf B51}, 3554 (1995).

\bibitem{Ral-Lud-Del-Buh}
D.C. Ralph, A.W.W. Ludwig, J. von Delft and R.A. Buhrman,
Phys. Rev. Lett. {\bf 72}, 1064 (1994).

\bibitem{Het-Kro-Her}
M.H. Hettler, J. Kroha, and S. Hershfield, Phys. Rev. Lett.
{\bf 73}, 1967 (1994).

\bibitem{Win-Alt-Mei}
N.S. Wingreen, B.L. Altshuler and Y. Meir, unpublished.

\bibitem{Mah}
G.D. Mahan, {\it Many-Particle Physics}, (Plenum, 1990).

\bibitem{Jan-etal}
A.G.M. Jansen, A.P. van Gelder, P. Wyder, and S. Strassler,
J. Phys. {\bf F11}, L15 (1981).

\end{references}
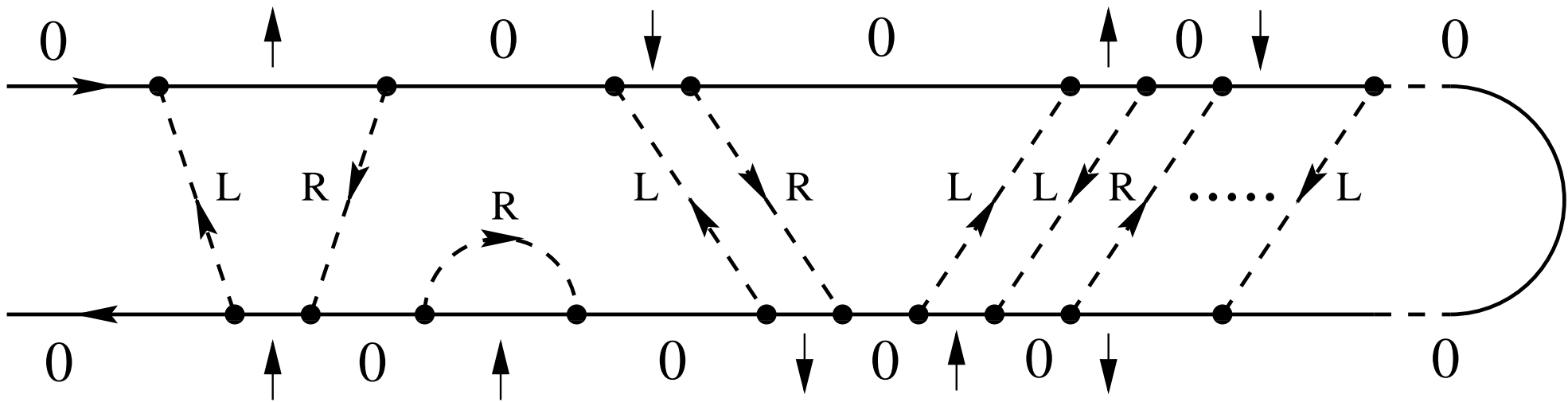
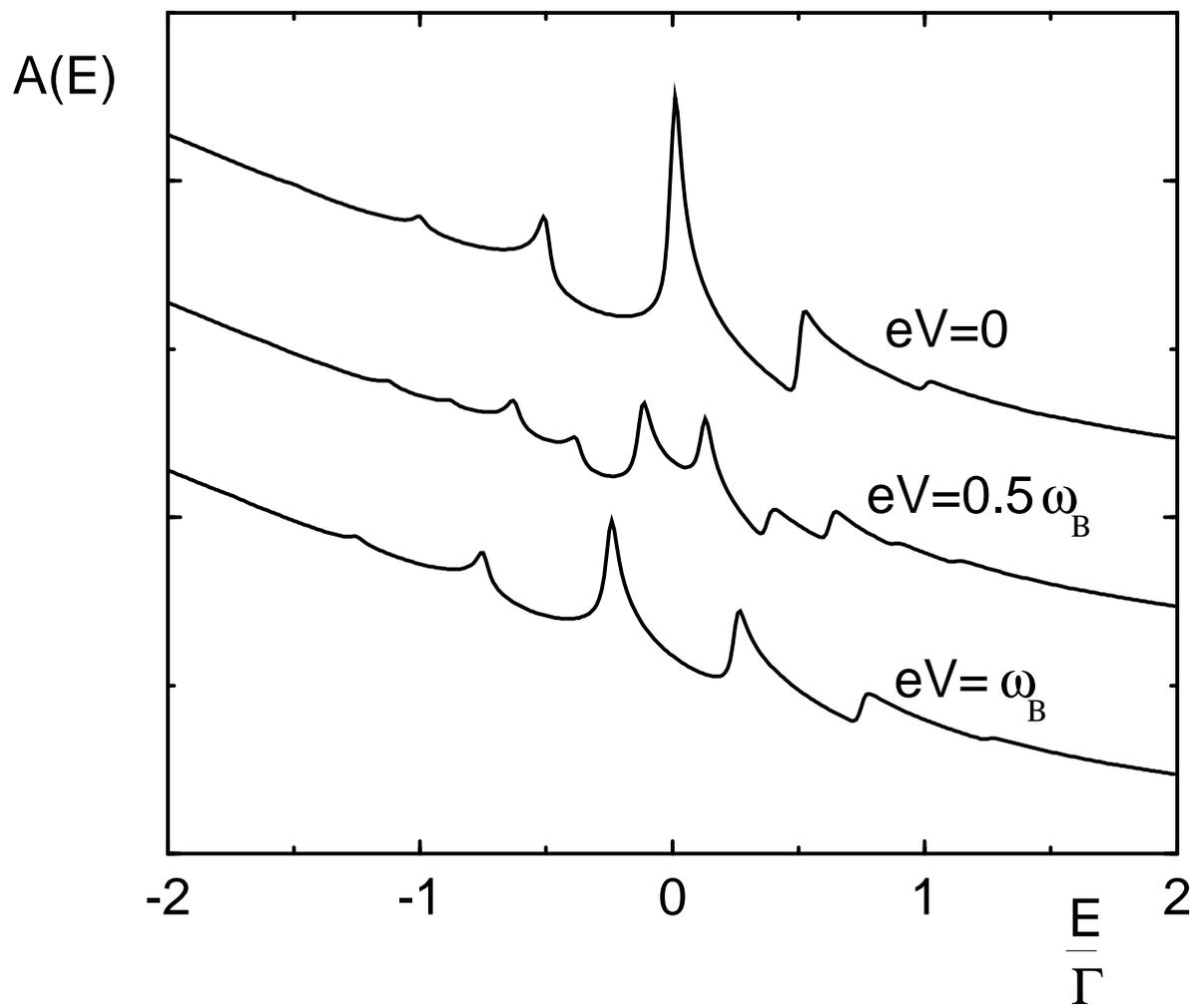
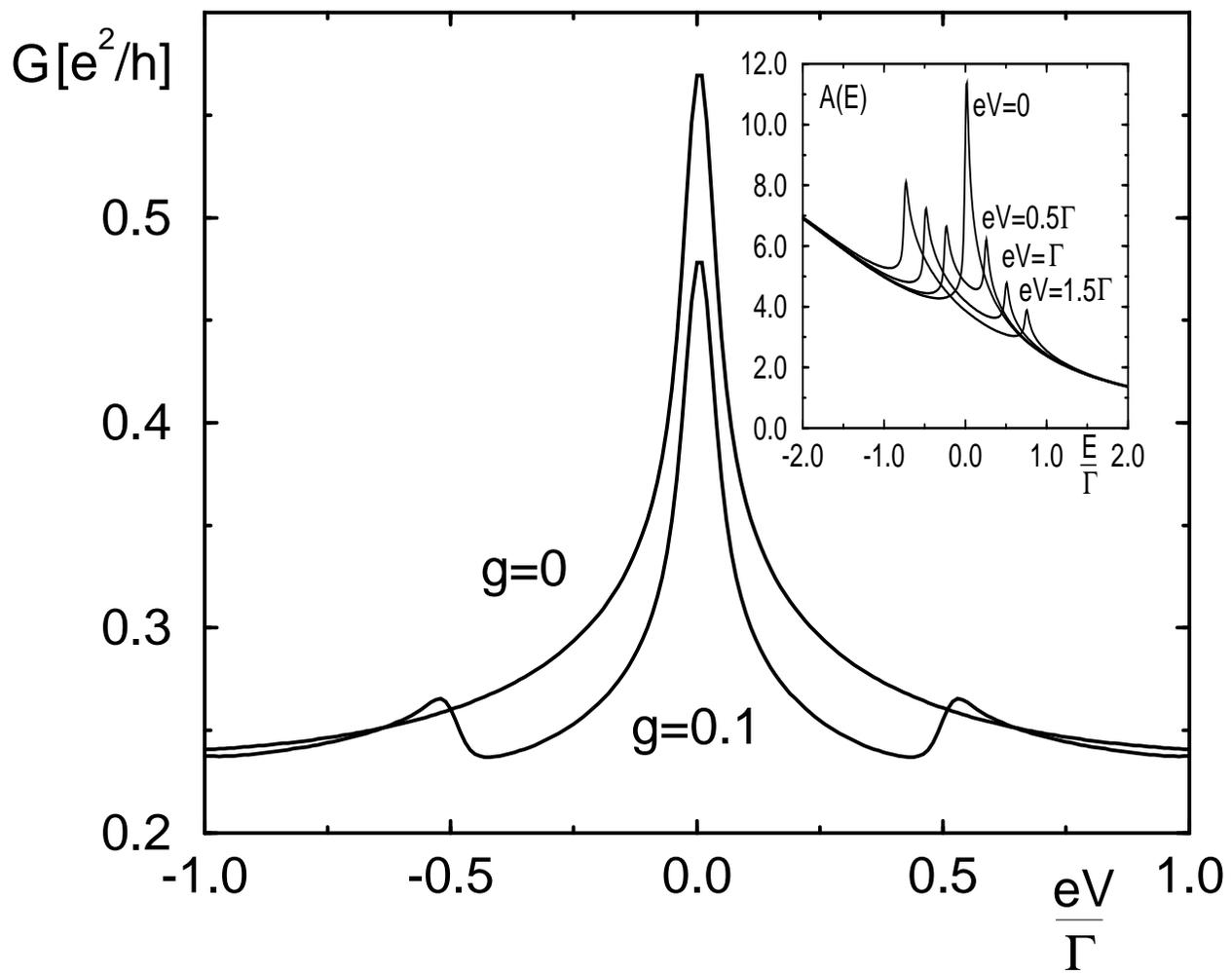
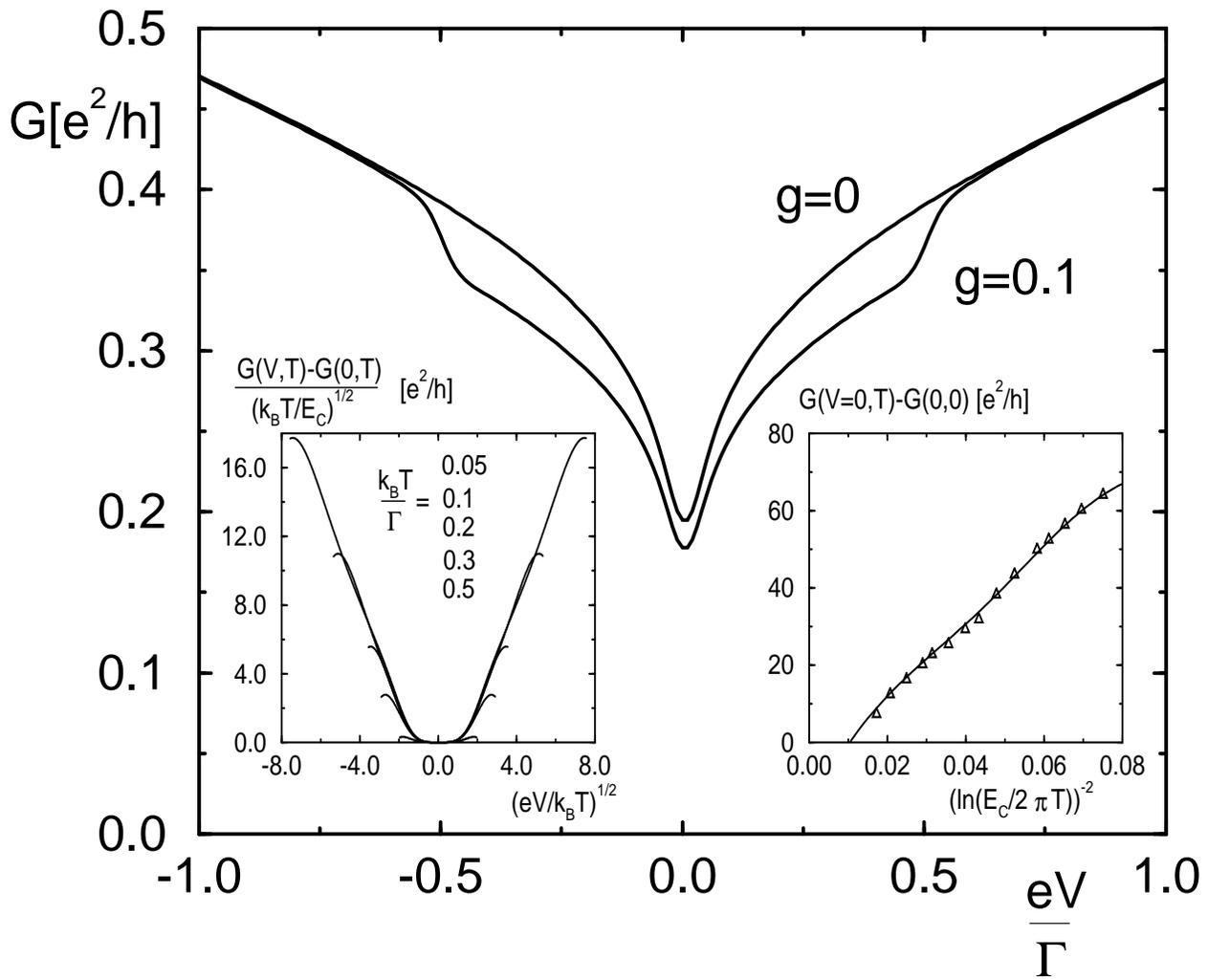
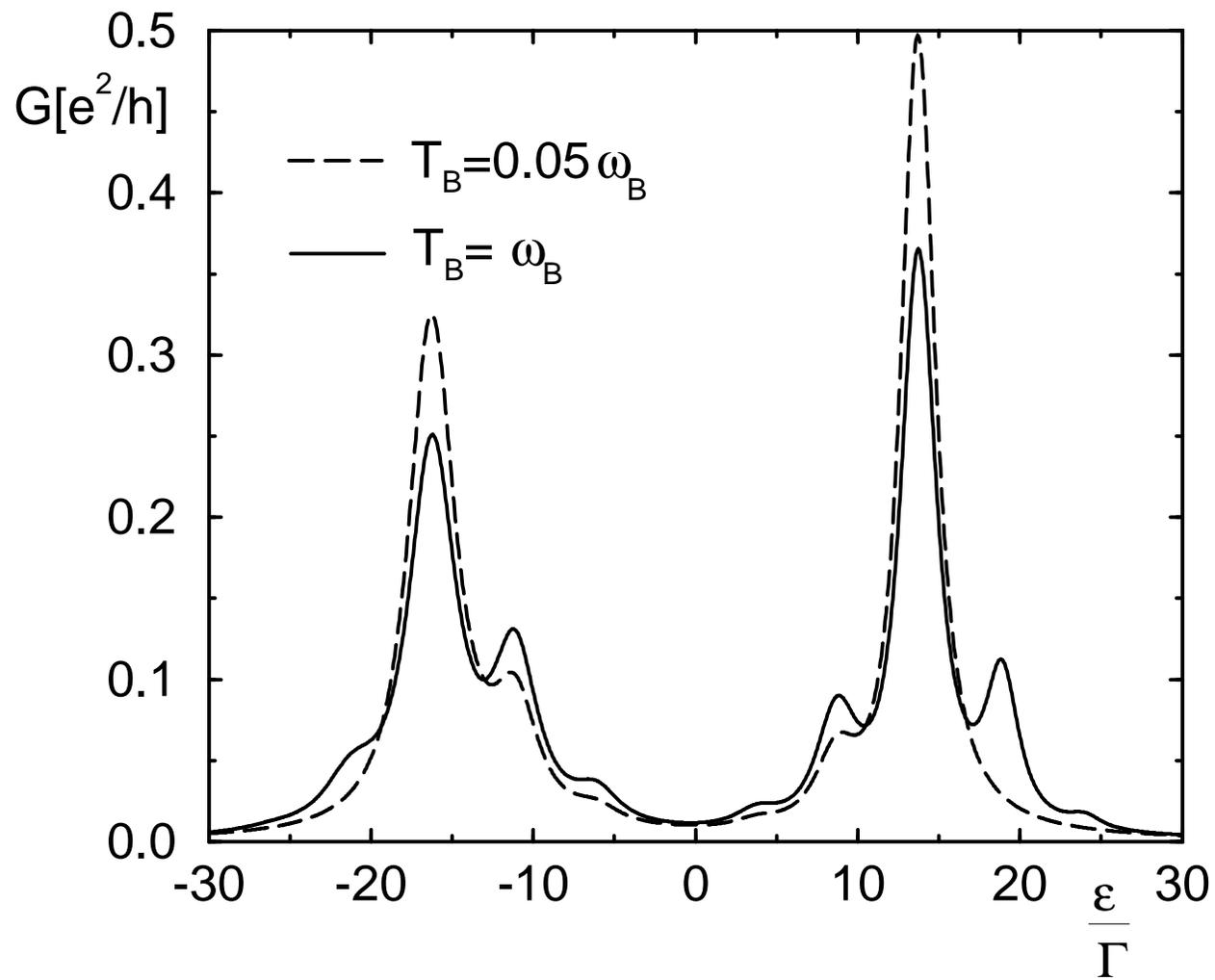
\end{document}